\documentclass[osa,twocolumn,showpacs,floatfix,bibnotes,amsmath]{revtex4}

\usepackage[dvips]{graphicx}

\usepackage{here}

\newcommand\pictc[5]{\begin{figure}
                       \centerline{
                       \includegraphics*[width=#1\columnwidth]{#3}}
                   \protect\caption{\protect\label{fig:#4} #5}
                    \end{figure}            }
\newcommand\pict[4][1.]{\pictc{#1}{!tb}{#2}{#3}{#4}}
\newcommand\rpict[1]{\ref{fig:#1}}

\newcommand\leqt[1]{\protect\label{eq:#1}}
\newcommand\reqtn[1]{\ref{eq:#1}}
\newcommand\reqt[1]{(\reqtn{#1})}

\newcounter{Fig}

\begin{document}
\begin{sloppy}

\title{Spatial optical solitons supported by mutual focusing}

\author{Jos\'e R. Salgueiro}
\author{Andrey A. Sukhorukov}
\author{Yuri S. Kivshar}

\affiliation{Nonlinear Physics Group, Research School of Physical
Sciences and Engineering, Australian National University, Canberra
ACT 0200, Australia}
\homepage{http://www.rsphysse.anu.edu.au/nonlinear}

\begin{abstract}
We study composite spatial optical solitons supported by two-wave
mutual focusing induced by cross-phase modulation in Kerr-like
nonlinear media. We find the families of both single- and two-hump
solitons and discuss their properties and stability. We also
reveal remarkable similarities between recently predicted
holographic solitons in photorefractive media and parametric
solitons in quadratic nonlinear crystals.
\end{abstract}


\maketitle

One of the standard physical mechanisms supporting the existence
of spatial optical solitons is the beam self-focusing due to
nonlinearity-induced change of the medium refractive index (see,
e.g., Ref.~\cite{Kivshar:2003:OpticalSolitons} and references
therein). In such a case, an optical beam induces an effective
waveguide and becomes self-guided in a nonlinear optical
medium.

A rather different physical mechanism is responsible for the
so-called mutual beam focusing which can be observed, for example,
in photorefractive two-wave mixing \cite{Vaupel:1997-1470:OL}. In
this case, a single beam diffracts because its self-induced
focusing is weak or negligible. However, the same beam can
demonstrate focusing in the presence of the other beam, due to the
mutual cross-phase-modulation interaction in a Kerr-like medium
with the third-order nonlinear susceptibility. As was predicted
recently, the effect of mutual focusing can be responsible for the
formation of a new kind of optical spatial soliton, the so-called
{\em holographic soliton} \cite{Cohen:2002-2031:OL}. Such a
soliton is created in the absence of the
self-action effect, solely due to the beam cross-coupling through
the induced Bragg reflection. Because the holographic soliton is described by two modes in the frames of the coupled-mode theory, it can be regarded as a special case of vector solitons, well studied in nonlinear optics.

The concept of a vector soliton was originally associated with
nonlinear interaction of two polarizations in a birefringent Kerr
medium. In this case, both self- and cross-phase modulation of the
beams produce a bound state of two orthogonally polarized field
components in the form of a self-trapped two-component {\em vector
soliton} \cite{Christodoulides:1988-53:OL}. The existence of
spatial vector solitons can be understood through the concept of
induced waveguiding. Indeed, when one of the soliton components
creates an effective waveguide in a nonlinear medium via the
self-modulation effect, the second component can be localized as a
guided mode of this waveguide, which creates a composite state in
the form of a vector soliton even beyond the bifurcation point~\cite{Haelterman:1993-1406:OL}. This
concept allows to introduce, in a simple and straightforward way,
both single- and multi-hump vector solitons, once it is accepted
that the soliton-induced waveguide can support higher-order modes~\cite{Kivshar:2003:OpticalSolitons}.

In the case of vector solitons due to mutual focusing the physics
of beam coupling cannot be grasped by the simple waveguiding
argument~\cite{Cohen:2002-2031:OL}. In particular, it is not
intuitively clear what would be the properties of such solitons
when the powers of the two constituents are significantly
dissimilar. Since the holographic solitons have been found only in
the special case when two optical beams have equal amplitudes (and
the corresponding mathematical problem reduces to a scalar
equation), we wonder if the mutual self-focusing may generate
stable composite solitons in a Kerr-like medium in a broad range
of experimental parameters, and if they can be compared with other
types of solitons studied earlier.

First of all, we note that mutual focusing is also known to
support parametric spatial solitons in {\em quadratic nonlinear
media}~\cite{Karamzin:1975-834:ZETF}. In general, such three-wave
parametric solitons consist of two coupled low-frequency beams and
the sum-frequency component. In this case, the phase-matching
energy-dependent interaction between the beams is {\em a key
physical mechanism for supporting the parametric solitons}. One of
the waves can be generated inside the crystal, however at least
two waves at the input are required to generate a three-wave
parametric soliton. This feature suggests that there may be a
strong similarity between different types of solitons supported by
mutual focusing.

In this Letter we analyze the composite spatial solitons supported
by mutual beam focusing in a Kerr-like nonlinear medium {\em in the
absence of the self-action effects}. We predict the existence of
continuous families of single- and two-hump composite solitons,
and discuss their stability and interaction.

In order to describe the mutual focusing of two optical beams in a
nonlinear medium with the third-order susceptibility, we consider
{\em the model for holographic solitons} recently introduced by
Cohen {\em et al.} \cite{Cohen:2002-2031:OL},
\begin{equation}   \leqt{model}
   \begin{array}{l} {\displaystyle
      i \frac{\partial u}{\partial z}
      + \frac{\partial^2 u}{\partial x^2}
      + \frac{ |v|^2 u }{1 + |u|^2 + |v|^2}
      = 0 ,
   } \\*[9pt] {\displaystyle
      i \frac{\partial v}{\partial z}
      + \frac{\partial^2 v}{\partial x^2}
      + \frac{ |u|^2 v }{1 + |u|^2 + |v|^2}
      = 0 ,
   } \end{array}
\end{equation}
where $u(x,z)$ and $v(x,z)$ are the normalized envelopes of the
two interacting waves in a waveguide geometry. The waves are
coupled only through the effective nonlinear cross-phase
modulation terms and the saturation effect is also taken into
account~\cite{Cohen:2002-2031:OL}. Since the mode coupling is 
effectively incoherent in this approximation, the partial mode powers, $P_1 = \int_{-\infty}^{+\infty} |u|^2 dx$ and $P_2 =
\int_{-\infty}^{+\infty} |v|^2 dx$, are conserved independently.

At low powers, nonlinear interaction is weak and the
diffraction-induced spreading of both beams is observed. However,
at higher powers mutual focusing of the beams becomes important.
Spatial solitons correspond to a balance between nonlinear and
diffraction effects; the corresponding solutions of
Eq.~\reqt{model} are sought in the form: $u(x,z) = U(x) \exp(i
\alpha_1 z + i \varphi_1)$ and $v(x,z) = V(x) \exp(i \alpha_2 z +
i \varphi_2)$, where $\alpha_{1,2}$ are the nonlinear propagation
constants and $\varphi_{1,2}$ are the arbitrary phases. The real
amplitude profiles $U(x)$, $V(x)$ satisfy the following system of
coupled equations,
\begin{equation}   \leqt{soliton}
   \begin{array}{l} {\displaystyle
      - \alpha_1 U
      + \frac{d^2 U}{d x^2}
      + \frac{ V^2 U }{1 + (U^2 + V^2)}
      = 0 ,
   } \\*[9pt] {\displaystyle
      - \alpha_2 V
      + \frac{d^2 V}{d x^2}
      + \frac{ U^2 V }{1 + (U^2 + V^2)}
      = 0 .
   } \end{array}
\end{equation}

We look for localized solutions describing bright solitons for
which light is localized in the central region,
$(U,V)(x\rightarrow \pm\infty) \rightarrow 0$. In a special
degenerate case when $\alpha_1 = \alpha_2$, the profiles of two
components coincide, $U(x) = V(x)$, and such solutions have been
recently described by Cohen {\em et
al.}~\cite{Cohen:2002-2031:OL}. In this Letter, we study a general
case when the amplitude profiles of two waves do not coincide.
Generally speaking, the two-wave soliton solutions can be found
only numerically, but the existence region of symmetric
single-hump solitons can be evaluated analytically. Indeed, from
the symmetry property for single-hump solitons, $\{U,V\}(x) =
\{U,V\}(-x)$,  it follows that the point  $x=0$ is a maximum, and
we use Eq.~\reqt{soliton} to obtain the corresponding inequalities
for the soliton peak amplitudes $U_m=U(0)$ and $V_m=V(0)$, $V_m^2
(1 + U^2_m + V_m^2)^{-1} > \alpha_1$ and $U_m^2 (1 + U_m^2 +
V_m^2)^{-1} > \alpha_2$. These two inequalities are compatible
only when $\alpha_{1,2}>0$ and $(\alpha_1+\alpha_2)<1$, so that
these conditions define the existence region of the single-hump
vector solitons. The soliton peak amplitudes are bounded from
below,
\begin{equation}   \leqt{max}
   U_m^2 > \frac{\alpha_2}{1 - (\alpha_1 + \alpha_2)},\quad
   V_m^2 > \frac{\alpha_1}{1 - (\alpha_1 + \alpha_2)} .
\end{equation}
Therefore, the peak intensity of one wave is proportional to the
propagation constant of the other component; this relation
reflects the fact that the two-wave soliton is supported solely by
the mutual focusing effect. Both intensities rapidly increase and
the saturation effect becomes important close to the boundary of the
existence region where $(\alpha_1+\alpha_2) \rightarrow 1$.

\pict{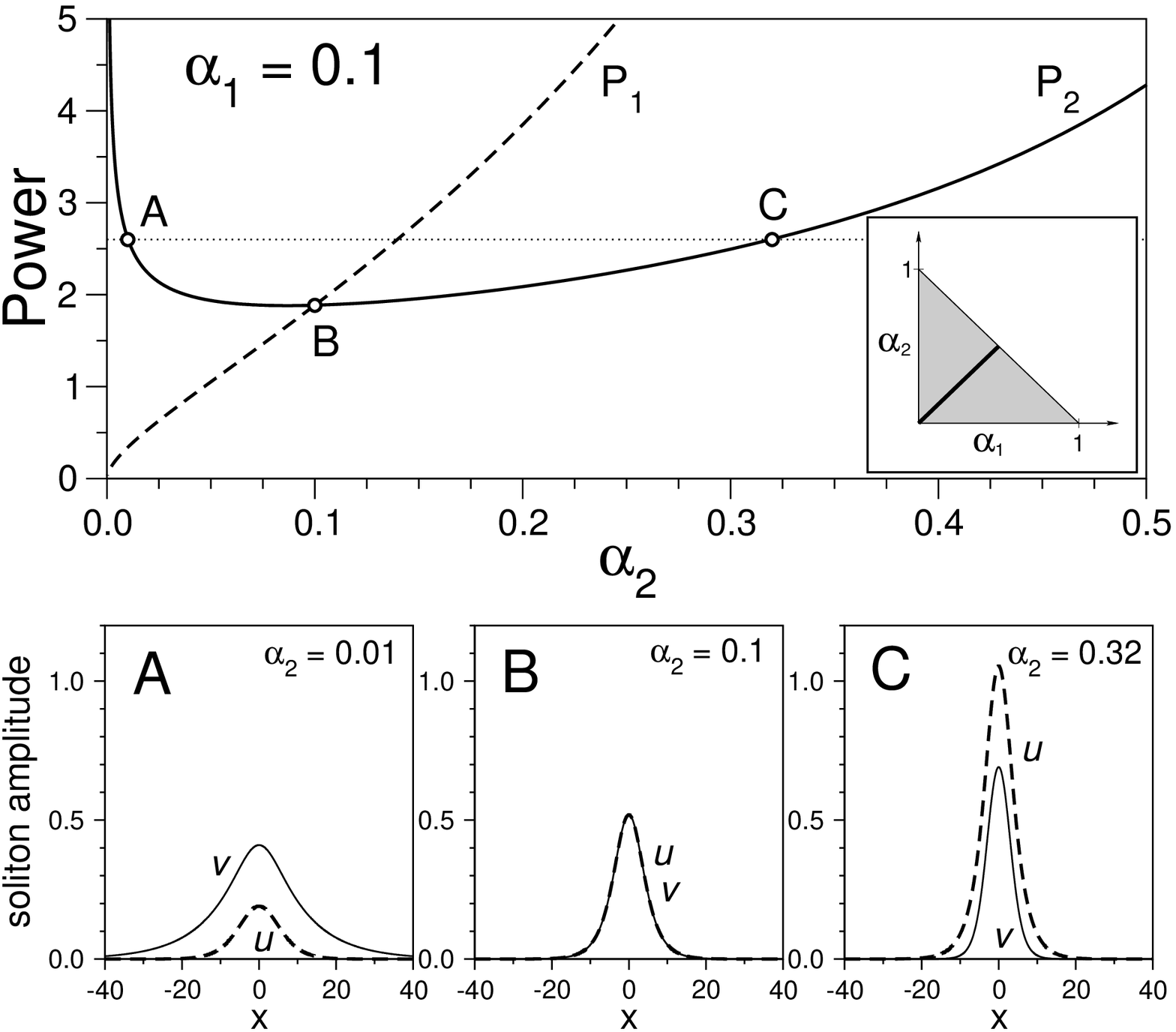}{power}{
Top: Powers $P_1$ and $P_2$ of the
soliton components as functions of the soliton parameter
$\alpha_2$ at fixed $\alpha_1=0.1$. Inset shows the existence
domain on the parameter plane, the solid line $\alpha_1=\alpha_2$
corresponds to the previously discussed
solitons~\cite{Cohen:2002-2031:OL}. Thin horizontal line indicates the
bistable solutions. 
Bottom: Examples A, B, and C of the two-component single-hump solitons supported by mutual focusing. Solid and dashed curves correspond to the fields $v$ and $u$, respectively. }

\pict{fig02.eps}{twoHump}{(a)-(d) Examples of two-component
two-hump solitons supported by mutual focusing for different vales
of $\alpha_2$ (indicated on the plots) and fixed $\alpha_1=0.5$. }

To find the soliton families, we solve Eq.~\reqt{soliton}
numerically and find localized solutions everywhere inside the
analytically estimated existence region on the parameter plane
$(\alpha_1,\alpha_2)$ shown in Fig.~\rpict{power}(top) as a
triangular domain in the inset. The characteristic dependencies of
the soliton partial powers are presented in
Fig.~\rpict{power}(top) as functions of the propagation constant
$\alpha_2$. The characteristic examples of the two-wave soliton
profiles are shown in Figs.~\rpict{power}(bottom), and they
correspond to the points A, B, and C marked on the power plot.  In
agreement with the analytical results~\reqt{max}, the amplitude
of the $u$ component increases for larger values of the
propagation constant $\alpha_2$, and the corresponding power
dependence $P_1(\alpha_2)$ is monotonic. On the other hand, the
power dependence $P_2(\alpha_2)$ for the $v$ component exhibits
bistability. The power increases for $\alpha_2\rightarrow 0$
because in the low-saturation regime the amplitude of the $v$
component is almost fixed by the value of $\alpha_1$, whereas the
beam width grows as $\sqrt{\alpha_2}$. In spite of such a
non-monotonic dependence of the power component, {\em the two-wave
solitons corresponding to both branches are stable}. The reason is
the following. In the system under consideration two partial
powers are conserved independently, as the mutual interaction is
incoherent. In this case, the soliton stability is defined by the
generalized Vakhitov-Kolokolov criterion for multi-component
solitons~\cite{Pelinovsky:2000-8668:PRE} which, as we have
verified for our model, predicts soliton stability. Thus,
single-hump vector solitons supported by mutual focusing are
stable.

We now search for {\em two-hump vector solitons} for which the
components $u$ and $v$ have symmetric unipolar and antisymmetric
bipolar profiles, respectively. We find that such solitons appear
only when $\alpha_1 < \alpha_2$. This is exactly one half of the
existence region for one-hump vector solitons in the inset shown
in Fig.~\rpict{power}(top). The characteristic examples of
two-hump solitons are presented in Fig.~\rpict{twoHump}, where we
show the modification of the soliton profiles when one of the
parameters is fixed. As the boundary of the existence domain is
approached, i.e. $\alpha_2 \rightarrow \alpha_1$, the separation
between two humps increases, and the two-hump solutions can be
considered as a bound state of two single-hump vector solitons.
However, for small $\alpha_2$ the unipolar component becomes much
wider, and it carries more power than the bipolar one. In this
limit the structure of two-hump vector solitons resembles that of
parametric solitons in nonlinear quadratic
media~\cite{Yew:1999-33:JNS}.

\pict{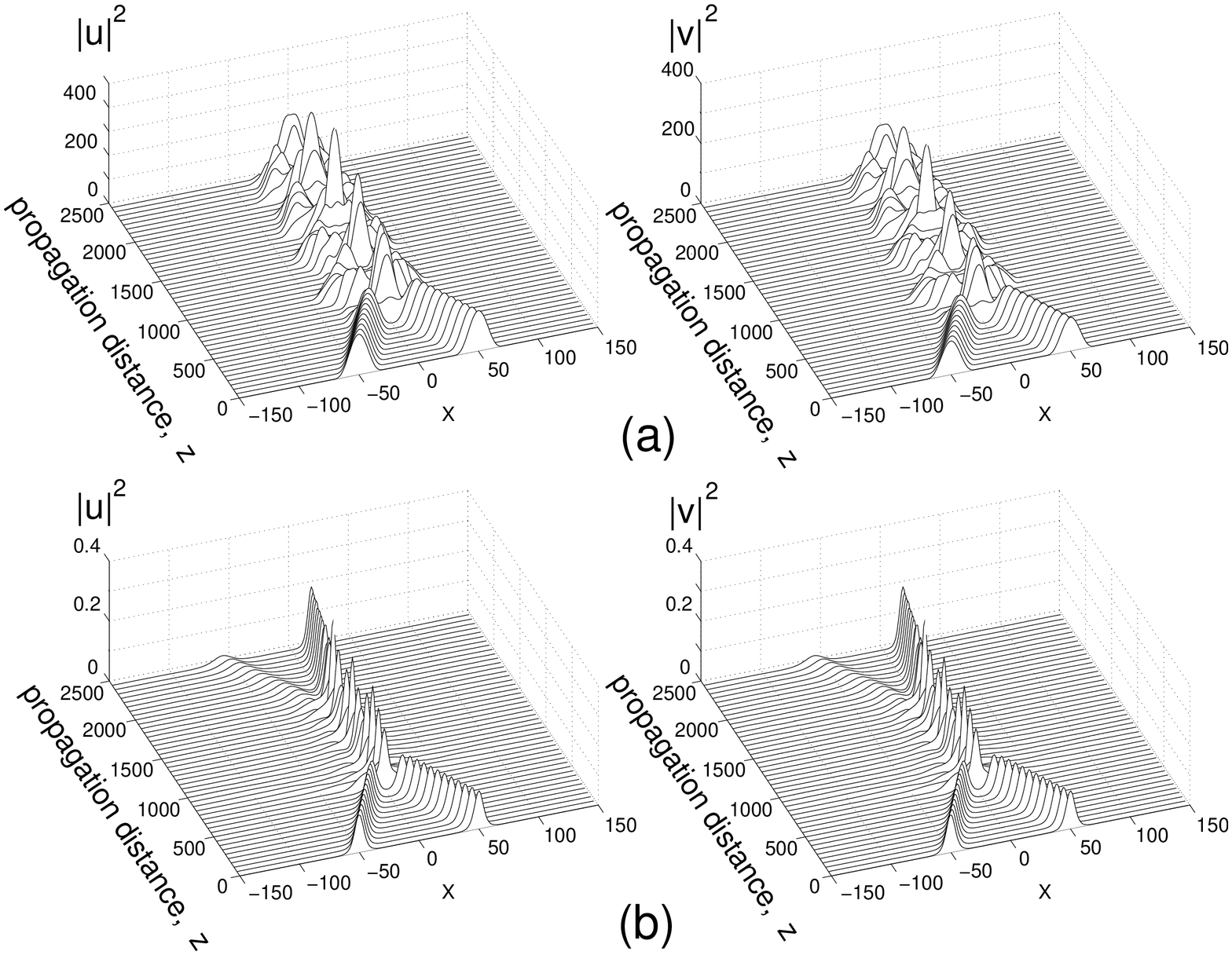}{bpm}{ 
Collision of two in-phase vector solitons
in (a)~high ($\alpha_1 = 0.39$, $\alpha_2 = 0.59$) and (b)~low
($\alpha_1 = 0.039$, $\alpha_2 = 0.059$) saturation regimes. }

Our results suggest that there should exist a link between
holographic and three-wave parametric solitons. Let us consider
the properties of parametric
solitons~\cite{Karamzin:1975-834:ZETF, Buryak:2002-63:PRP} when the phase mismatch
$\Delta k$ is large. Then, the generated sum-frequency wave has a
small amplitude, which can be found using the cascading
approximation, $E_3 \sim E_1 E_2 / \Delta k$. After substituting this expression into the governing
equations for the low-frequency beams, we obtain the model which
is mathematically equivalent to Eqs.~\reqt{model} in the
low-saturation limit. Therefore, many similarities between
holographic and parametric solitons are not a coincidence, but
rather a manifestation of the similar physical origin of such
solitons~-- the effect of mutual focusing.

We would like to stress that, despite some similarities, the
soliton dynamics in the low- and high-saturation regimes can be
substantially different, as is clearly illustrated in
Figs.~\rpict{bpm}(a,b) for the soliton collisions. In both cases
solitons merge, however only in the high-saturation regime strong
oscillations develop, see Fig.~\rpict{bpm}(a). This happens
because, in the regions with high beam intensity, the medium
response is almost linear, and persistent beating between several
modes can develop. On the other hand, in the low-saturation case
oscillations quickly decay due to strong radiation, see
Fig.~\rpict{bpm}(b).

In conclusion, we have analyzed two-component spatial solitons
supported by mutual focusing of two optical beams in the absence
of the self-phase modulation effects. One of the possible
realization of these solitons is through the holographic focusing
effect recently discussed by Cohen {\em et al.}
\cite{Cohen:2002-2031:OL}. We have demonstrated the existence of
the continuous families of single- and two-hump vector solitons,
and have discussed their stability. Additionally, we have
emphasized a key difference between the formation of the
convectional vector solitons and the two-component solitons
supported by mutual focusing, but also revealed their links to the
parametric quadratic solitons.

We thank E.~A.~Ostrovskaya and D.~E. Pelinovsky for useful
discussions. The work was partially supported by the Australian
Research Council. J. R. Salgueiro acknowledges a postdoctoral
fellowship conceded by the Secretar\'{\i}a de Estado de
Educaci\'on y Universidades of Spain and supported by the European
Social Fund.

\end{sloppy}
\end{document}